\documentclass[a4paper,fleqn]{cas-dc}

\usepackage[compress]{natbib}

\usepackage{pifont}
\usepackage{caption}
\usepackage{subfig}

\def\tsc#1{\csdef{#1}{\textsc{\lowercase{#1}}\xspace}}
\tsc{WGM}
\tsc{QE}
\tsc{EP}
\tsc{PMS}
\tsc{BEC}
\tsc{DE}

\begin{document}
\let\WriteBookmarks\relax
\def\floatpagepagefraction{1}
\def\textpagefraction{.001}

\shorttitle{Real-time Instance Segmentation of Surgical Instruments}

\shortauthors{J.C. Angeles-Ceron et~al.}

\title [mode = title]{Real-time Instance Segmentation of Surgical Instruments using Attention and Multi-scale Feature Fusion}        

\author[1]{Juan Carlos Ángeles-Cerón}
                     
\author[1]{Gilberto Ochoa-Ruiz}[type=author,
                        auid=000,bioid=1,                     orcid=0000-0002-9896-8727]\ead{gilberto.ochoa@tec.mx}
                        
\author[1]{Leonardo Chang}
\author[2,3]{Sharib Ali}[type=author,
                        auid=000,bioid=1,                     orcid=0000-0003-1313-3542]\ead{sharib.ali@eng.ox.ac.uk}

\cortext[cor1]{Corresponding authors}

%
\address[1]{Tecnologico de Monterrey, Escuela de Ingeniería y Ciencias, México}
\address[2]{Institute of Biomedical Engineering (IBME), Department of Engineering Science, University of Oxford, Oxford, UK}
\address[3]{Oxford NIHR Biomedical Research Centre, University of Oxford, Oxford, UK}

\begin{abstract}
Precise instrument segmentation aid surgeons to navigate the body more easily and increase patient safety. While accurate tracking of surgical instruments in real-time plays a crucial role in minimally invasive computer-assisted surgeries, it is a challenging task to achieve, mainly due to 1) complex surgical environment, and 2) model design with both optimal accuracy and speed. Deep learning gives us the opportunity to learn complex environment from large surgery scene environments and placements of these instruments in real world scenarios. The Robust Medical Instrument Segmentation 2019 challenge (ROBUST-MIS) provides more than 10,000 frames with surgical tools in different clinical settings. In this paper, we use a light-weight single stage instance segmentation model complemented with a convolutional block attention module for achieving both faster and accurate inference. We further improve accuracy through data augmentation and optimal anchor localisation strategies. To our knowledge, this is the first work that explicitly focuses on both real-time performance and improved accuracy. Our approach out-performed top team performances in the ROBUST-MIS challenge with over 44\% improvement on area-based multi-instance dice metric MI\_DSC and 39\% on distance-based multi-instance normalized surface dice MI\_NSD. We also demonstrate real-time performance ($> 60$ frames-per-second) with different but competitive variants of our final approach.
\end{abstract}


\begin{keywords}
Deep learning\\ MIS instance segmentation\\ Real-time instance segmentation\\ Single-stage segmentation\\ Attention Modules\\ Multi-scale feature fusion 
\end{keywords}

\maketitle

\section{Introduction}
\label{sec:introduction}

%
%
%
Surgical site infection (SSI) has been the most common cause of hospital-acquired infection and the most common way of infection transmission in patients undergoing surgery~\cite{Giorgio2004:JAMASurg,Daniel2019:JAMANet}. It is therefore imminently important to develop strategies for reducing such infection rates. Minimally invasive surgical (MIS) procedures compared to open surgery lowers such risks ~\cite{Daniel2019:JAMANet}. For these reasons and due to the growth the data science in the operating room applications, there exists an increasing demand of computer assisted/robotic surgery to improve the efficacy of MIS~\cite{Bartoli_2012, Sheetz2019:JAMANet}.

Computer-assisted minimally invasive surgery such as endoscopy has grown in popularity in recent years. However, due to the nature of these procedures, issues like limited field-of-view, extreme lighting conditions, lack of depth information and difficulty in manipulating operating instruments demand strenuous amounts of effort from the surgeons~\cite{Ross_2021}. Surgical data science applications \cite{maierhein2021surgical} could provide physicians with context-aware assistance during minimally invasive surgery in order to overcome these limitations and increase patient safety.

One of the main forms of assistance is providing accurate tracking of medical instruments using computer vision techniques, such as object detection and localisation or instance segmentation methods \cite{Ward2021}. These systems are expected to be a crucial component in tasks ranging from surgical navigation, skill analysis and complication prediction during surgeries, as well as other computer integrated surgery (CIS) applications \cite{Maierhein_2021, fu2021}. Nonetheless, methods for accurate tracking of instruments are often deployed in difficult operational scenarios in which the presence of bleeding, over or under exposed frames, smoke, reflection and other types of artifacts are oftentimes unavoidable \cite{Bodenstedt_2018}. The net effect of these issues increases the missed detection rates in endoscopic surveillance and in general, it limits the overall robustness of computer vision algorithms, hampering the adoption of AI-based tools in this context \cite{Ali_2021}. Moreover, real-time deployment of such tools is of tremendous value and one of the major requirement for it to be applied in clinical setting. 
Therefore, the development of robust and real-time techniques that can be effectively deployed during real endoscopic interventions is of utmost importance. In this regard, recent years have seen a significant increase in the number of computer vision contests geared towards endoscopy. More specifically, the Robust Medical Instrument Segmentation (ROBUST-MIS) Challenge \cite{Ross_2021} at the International Conference on Medical Image Computing and Computer Assisted Interventions (MICCAI) has sought to address some of the issues discussed above. This challenge represents an important and necessary effort to  encourage the development of robust models for surgical instrument segmentation, integrating the developments in computer-assisted surgeries, and benchmark the generalization capabilities of the developed methods on different clinical scenarios. 

Furthermore, the challenge organizers provide a large high-quality dataset in an effort to overcome one of the major bottlenecks of the development of robust methodologies, i.e., lack of annotated multi-instance instrument segmentation data. The development of surgical tool navigation and tracking methods in a complex environment will enable improved patient care during surgery by maximising the focus of surgeons. It will also accelerate research in the direction of robotic surgery. 

Previous approaches for instance segmentation submitted to the ROBUST-MIS challenge have been mostly based on two-stage detectors such as Mask R-CNN \cite{He_2018}. While these models exhibited decent performances in terms of robustness, they suffer from high inference times due to well-known architectural limitations of such models, preventing them for achieving real-time performance (i.e. mean average test time of only around 5 frames-per-second, fps). However, real-time performance is mandatory in order to fully exploit the capabilities of tracking applications in surgeries. While deep learning methods using lightweight models are also available, they fail to robustly segment objects in endoscopy imaging~\cite{Ali2020}. Similarly, there is a trend in using ensemble models to improve the overall segmentation accuracy in these images. However, combining few models together drastically increases the inference time, thereby making it less feasible to deploy such models in clinical settings~\cite{xu2020}.

In order to overcome the current inference limitations, while maintaining a robust performance in terms of tool segmentation results, we propose a new approach based on the single-stage model for instance segmentation. Although recent years have seen an steady increase in the search for more capable one stage detectors and instance segmentation architectures (for example PolyYolo~\cite{polyyolo}, BlendMask~\cite{blendmask} and Solov2~\cite{solov2}), the majority of these models have not been used in the context of endoscopic computer vision due to the performance gap they still present. The  YOLACT++ \cite{Bolya_2020} architecture is one of the most recent methods for real-time instance segmentation and it is particularly appealing due to its simplified architecture capable of learning to localize instance masks automatically with minimum computational overhead. It does so by generating a dictionary of non-local prototype masks over the entire image and predicting a set of linear combination coefficients per instance. Thus, for this contribution we have used YOLACT++ as a baseline architecture upon which we have developed several improvements to make it suitable for robust surgical instrument instance segmentation. To this extend, we have explored the use of attention modules on the outputs of the network's backbone and feature pyramid network (FPN) at multiple scales. Moreover, we have additionally carried out a series of optimization techniques by analysing the worst-performing frames of the best model in our experiments. The optimization techniques includes stronger data augmentation, anchor optimization, and multi-scale feature fusion. 

Our main contributions include:
\begin{itemize}
    \item A real-time single-stage instance segmentation framework with attention mechanisms 
    \item Use of domain-targeted data augmentation
    \item Anchor box optimization via a differential evolution search algorithm~\cite{Zlocha_2019} 
    \item Exploiting global contextual features by integrating multi-scale fusion in the network's backbone
    \item Thorough analysis of the worst-case samples to evaluate each of the tested model
\end{itemize}



The rest of the paper is organized as follows. In Section \ref{sec:related_work} we present previously published work related to medical instrument segmentation, instance segmentation methods, attention mechanisms and multi-scale feature fusion network. In Section~\ref{sec:materials_and_method}, we present the details of the ROBUST-MIS dataset and our proposed approach for surgical instrument segmentation. Section \ref{sec:resultsandExperiments} presents our data preparation, experimental setup and results. Next, in Section \ref{sec:discussion} we discuss the effects of the different types of network configurations and future directions. Finally, Section \ref{sec:conclusion} concludes the paper.
\section{Related work}
\label{sec:related_work}
In this section, we will discuss some of the most important aspects to understand the proposed contribution, namely: instance segmentation and its current limitations, recent works in attention mechanisms, anchor box optimization techniques specifically tailored to the addressed problem and finally, multi-scale fusion networks that followed to make our extended instance segmentation model more robust.

\subsection{Deep learning for instrument segmentation}
Deep learning has accelerated research for surgical instrument segmentation and the public access of labelled data via instrument segmentation challenges mostly at EndoVis (refer to \cite{Allan20192017RI},~\cite{Allan20202018RS},~\cite{Ross_2021}) have contributed to these developments over recent years. Built upon the UNet~\cite{Ronneberger_2015}, LinkNet and TernausNet was developed for instrument segmentation on robotic surgery dataset~\cite{shvets2018}, acquired by da Vinci Xi surgical system of several different porcine procedures made available in EndoVis17~\cite{Allan20192017RI}. \cite{milletari2018} proposed a convolutional long short term memory (LSTM) with deep residual networks in coarse-to-fine strategy showing greater improvements over other state-of-the-art approaches including UNet~\cite{Ronneberger_2015} and FCN~\cite{Long:CVPR15} on EndoVis 2015 instrument segmentation challenge dataset{\footnote{https://endovissub-instrument.grand-challenge.org/EndoVisSub-Instrument/}} focused on laparoscopic and robotic surgery. 

\newpage

The recent “Robotic Instrument Segmentation Sub-Challenge” introduced also at EndoVis was oriented towards binary segmentation, sub-component segmentation and instrument identification and instance segmentation tasks~\cite{Ross_2021}. The challenge was focused to assess the robustness and generalization capabilities of the deep learning models. Most of the competing methods in this challenge have been mostly based on Mask-RCNN \cite{mask-rcnn} implementations and its variants. Participants also explored methods such as OR-UNet \cite{orunet}, DeepLabV3+ \cite{deeplab}, U-Net \cite{unet} and RASNet \cite{rasnet}. The best performing methods for the binary segmentation task were OR-UNet and DeepLabv3+ with pre-trained ImageNet encoders. Some other contestants also explored the use of ensemble methods, but they were typically limited in speed, and thus some of the most robust methods are incapable of attaining the real-time performances required for realistically segmenting and tracking objects on endoscopic video data. A double decoder-encoder network was explored for faster binary mask segmentation on ROBUST-MIS'19 daataset~\cite{Jha:BHI2021} that outperformed several state-of-the-art methods. Recently, a one-shot instrument segmentation method~\cite{ZHAO2021102240} using anchor guided meta-learning approach was proposed and validated on several publicly available datasets.  

\subsection{Real-time instance segmentation methods}
While extensive research has been conducted for the development of real-time object detection and semantic segmentation  models, few works have tackled the problem of real-time instance segmentation~\cite{mask-rcnn,yolact}. This is due to the increased complexity in the instance segmentation task that requires prediction of instance labels and pixel-level segmentation simultaneously. 
One-stage methods~\cite{polyyolo,center-mask, blendmask} though conceptually faster than two-stage methods (e.g., Mask RCNN~\cite{mask-rcnn}), still require many non-trivial computations (e.g., mask voting). This severely limits their speed making them not unsuitable for real-time applications. In contrast, recent methods~\cite{yolact,Bolya_2020} make use of lightweight assembly of masks (only a linear combination is used), making the approach very efficient. Although YOLACT~\cite{yolact} was one of the first real-time one-stage instance segmentation approach, the accuracy gap compared to Mask R-CNN \cite{mask-rcnn} was still significant. While Mask R-CNN is based on a two-stage object detector (e.g., Faster R-CNN \cite{faster-rcnn}), YOLACT \cite{yolact} is built on one-stage detector (RetinaNet \cite{retina-net}) that directly predicts boxes without proposal step, limiting its accuracy. This was partially addressed with the introduction of Yolact++~\cite{Bolya_2020} incorporates deformable convolutions into the backbone network, improving the feature sampling and yielding an improved accuracy. Furthermore, the prediction head was optimized with better anchor scale and aspect ratio choices for an increased object recall. 
\subsection{Attention mechanisms}
Attention has been able to boost model performance across a wide range of computer vision tasks, such as image captioning \cite{You_2016}, visual question answering \cite{Xu_2016}, and visual attribute prediction \cite{Seo_2018}. Attention allows the network to focus on the most relevant features without the need of additional supervision, preventing redundant use of information and extracting salient features that are useful for a given task. Attention mechanisms enable convolutional neural networks to overcome the size limitations of its receptive field, as it has proven to be excellent at extracting global dependencies between inputs and outputs, thus, improving the modeling of long-range dependencies even at opposite ends of and image \cite{Sinha_2021}. For example, similar textures may appear in different parts of an image, several disjoint semantic cues may provide insight to the general classification of an image, and an object might present complex and occluded parts throughout an image \cite{Chaudhari2019AnAS}. In the context of medical instance segmentation, multiple attention-based models \cite{Kaul_2019, Gu_2020, Sinha_2021} have obtained state-of-the-art performance in fields like brain tumor, skin cancer and lung lesion segmentation on CT scans and X-rays. Until now, instance segmentation of medical instruments in laparoscopic surgeries using attention mechanisms has not been fully explored.

\subsection{Multi-scale feature fusion}
Due to the wide range of scale variation of objects found in instance segmentation, multi-scale features are essential for robust performance \cite{Wang_2020}. Therefore, multi-scale feature aggregation is an adequate strategy to create detailed parsing maps \cite{Ding_2018}. Current methods address this issue by using encoder-decoder architectures \cite{Ronneberger_2015, Milan_2017} that combine high level and low level features at a single scale \cite{Long_2015} or multiple scales \cite{Ronneberger_2015}. However, these approaches suffer from redundant use of information \cite{Sinha_2021}. In the field of medical instrument segmentation, the use of multi-scale aggregation has not been fully explored, especially combined with attention for improved robustness.

\section{Materials and Method}
\label{sec:materials_and_method}
In this section we present details on the dataset used in our study and describe our proposed framework for multi-instance surgical instrument segmentation. 
\subsection{The ROBUST-MIS challenge dataset}
\label{sec:robust_mis}

\begin{table}[t!]
\fontsize{7}{8}\selectfont
\centering

\caption{Training and test sample distribution for each both training and test stages of the ROUBUST-MIS challenge~\cite{Ross_2021}. The quantities in parenthesis represent the \% of frames with no instrument instance.}\label{tab:data_distribution}
\begin{tabular}{l|l|lll}
\toprule
\multirow{2}{*}{\textbf{Procedure}}
& \multirow{2}{*}{\textbf{Training}}    & \multicolumn{3}{c}{\textbf{Testing}}       \\ 
 & \multicolumn{1}{l}{} & \multicolumn{1}{|l}{Stage 1}              & \multicolumn{1}{l}{Stage 2}           & \multicolumn{1}{l}{Stage 3}   \\ \hline
{{\begin{tabular}[l]{@{}c@{}}Procto- \\ colectomy \end{tabular}}}                        & 2,943 (2\%)      & 325 (11\%)       & 255 (11\%)       & 0                    \\
{{\begin{tabular}[l]{@{}c@{}}Rectal\\ resection \end{tabular}}}                          & 3,040 (20\%)     & 338 (20\%)       & 289 (15\%)       & 0                    \\
{{\begin{tabular}[l]{@{}c@{}}Sigmoid\\  resection*  \end{tabular}}}  
                    & 0                    & 0                    & 0                    & {{\begin{tabular}[l]{@{}c@{}} 2,880 \\(23\%) \end{tabular}}}      \\ \hline
TOTAL                                  & 5,983 (17\%)     & 663 (15\%)       & 514 (13\%)       & {{\begin{tabular}[l]{@{}c@{}}2,880 (23\%) \end{tabular}}}    \\ \hline
\multicolumn{5}{l}{*Unknown surgery} 
\end{tabular}
\end{table}
For our experiments we made use of the Robust Medical Instrument Segmentation (ROBUST-MIS) \cite{Maierhein_2021} challenge dataset which is the first large-scale annotated MIS dataset. The dataset is comprised of a total of 10,040 annotated video frames extracted from 30 minimally invasive daily-routine surgical procedures and includes detailed segmentation ground truth masks for the surgical instruments present in these frames. 

\begin{figure}[t!]
    \centering
    \includegraphics[width=0.98\linewidth]{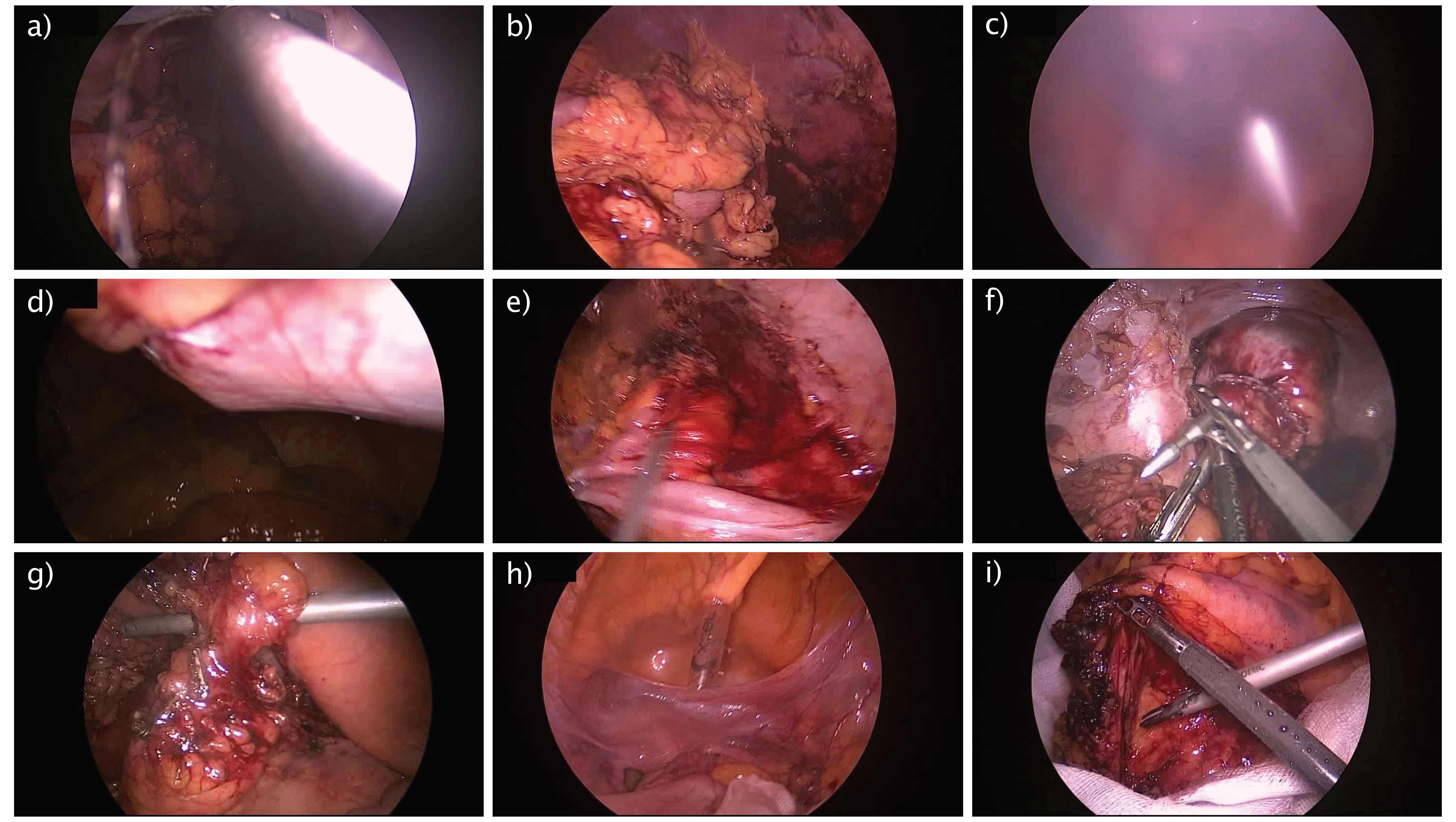}
    \caption{\textbf{Challenging images present in test dataset.} Stage 3 test data consisted of several frames that consisted of a) instrument flare, b) parital occlusions due to blood, c) occlusion due to smoke, d) underexposed regions with instrument, e) motion blur, f) multiple different instruemnts in the scene, g) partial occlusion due to organ, h) transparent instrument and i) different instruments crossing.}
    \label{fig:sampleimages}
\end{figure}

The surgical procedures include 10 rectal resection procedures, 10 proctocolectomy procedures, and 10 sigmoid resection procedures. The image resolution of all the provided frames is $960 \times 540$ pixels. In order to measure the robustness test, the dataset is comprised of three unique test sets and divided into different stages:
\begin{enumerate}
\item[] \textbf{Stage 1:} Test data taken from the same procedures from which the training data were extracted\\
\item[]\textbf{Stage 2:} Test data taken from the exact same type of surgery as the training data but from procedures (patients) not included in the training\\
\item[] \textbf{Stage 3:} Test data taken from a different (unseen) but similar type of surgery and different (unseen) patients
\end{enumerate}
The detailed training and test distribution is summarised in Table \ref{tab:data_distribution}. However, for testing we have used only test set 3, \textit{i.e. Stage 3}, which is from an unseen Sigmoid resection procedure and allows us to validate on the generalisability of the proposed framework directly. Sample images for challenging frames provided in the test set 3 is shown in Figure~\ref{fig:sampleimages}.
%
\subsection{Method}

%

\begin{figure*}[t!]
    \centering
    \includegraphics[width=\textwidth]{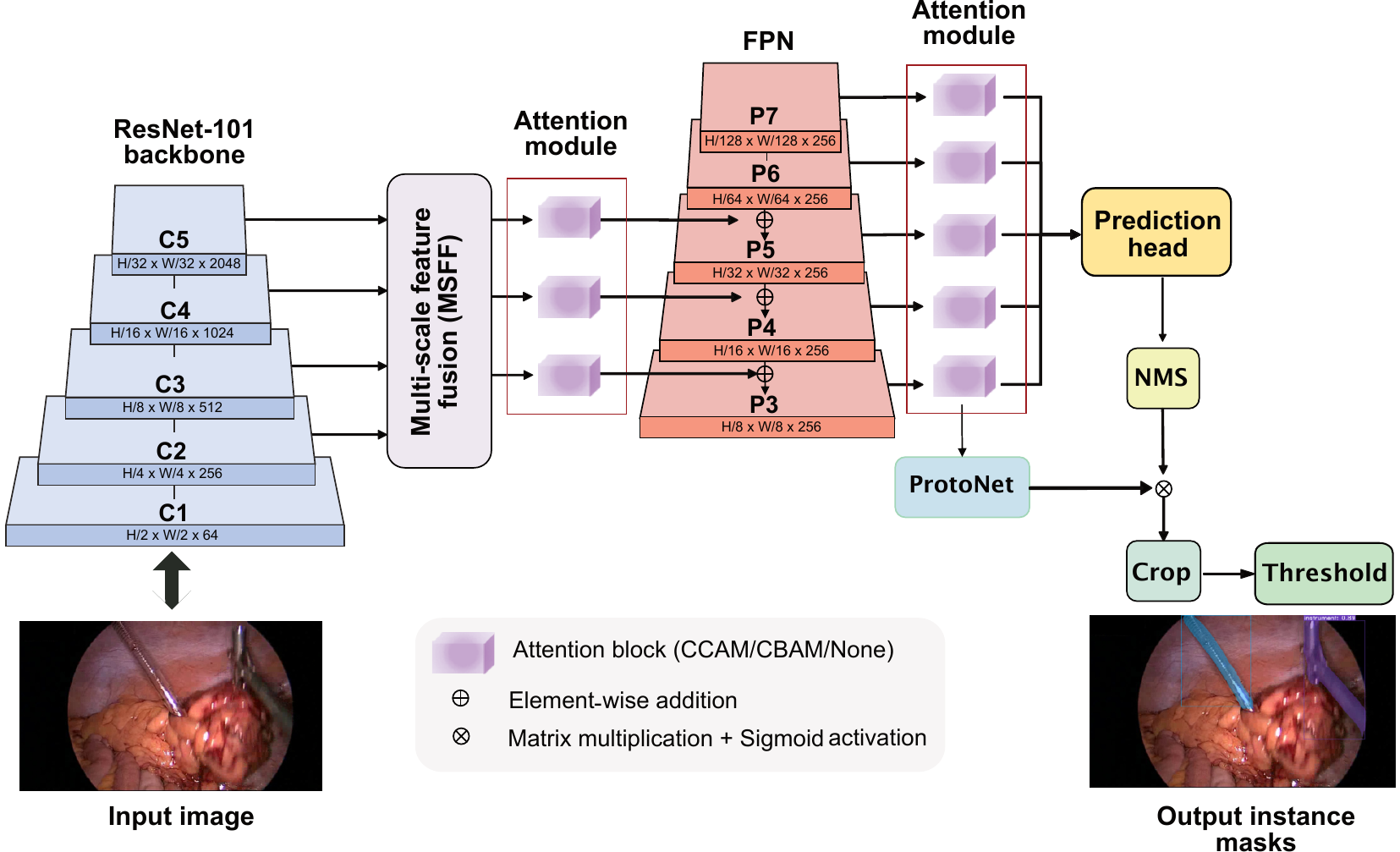}
    \caption{
     \textbf{Overview of our proposed framework.} Our architecture is built upon popular single-stage YOLACT++ and comprises of additional multi-scale feature fusion and attention modules. It is to be noted that the attention modules is easily interchangeable from criss-cross attention (CCAM) to convolutional block attention module (CBAM) or none.}
    \label{fig:architecture}
\end{figure*}


This section describes our proposed approach for multi-instance segmentation of medical instruments in the ROBUST-MIS dataset. Our proposed framework (see Figure \ref{fig:architecture}) is built upon the single-stage YOLACT++~\cite{Bolya_2020} instance segmentation architecture. For our framework we employ ResNet-101~\cite{He_2015} as the backbone network, followed by a multi-scale feature fusion (MSFF) module used to aggregate contextual information from the feature maps across all scales (i.e. high-to-low resolution feature representations). Each of these contextually rich fused features are then passed through attention modules to further refine these representations before being forwarded to the feature pyramid network (FPN)~\cite{Lin_2017}. A second set of attention modules are then applied to further enhance the FPN output features allowing for an improved performance of the prototype network and our anchor-optimized prediction head. Finally, we perform classical non-maximum suppression for final mask instance prediction which is then combined with the prototype mask and the cropping provides the predicted bounding box. 

Below we describe our MSFF module, attention mechanisms, and the anchor optimization used our framework. 

\subsubsection{Multi-scale feature fusion}
\label{sec:ms_fusion}
\begin{figure}[t!]
    \centering
    \includegraphics[width=0.98\linewidth]{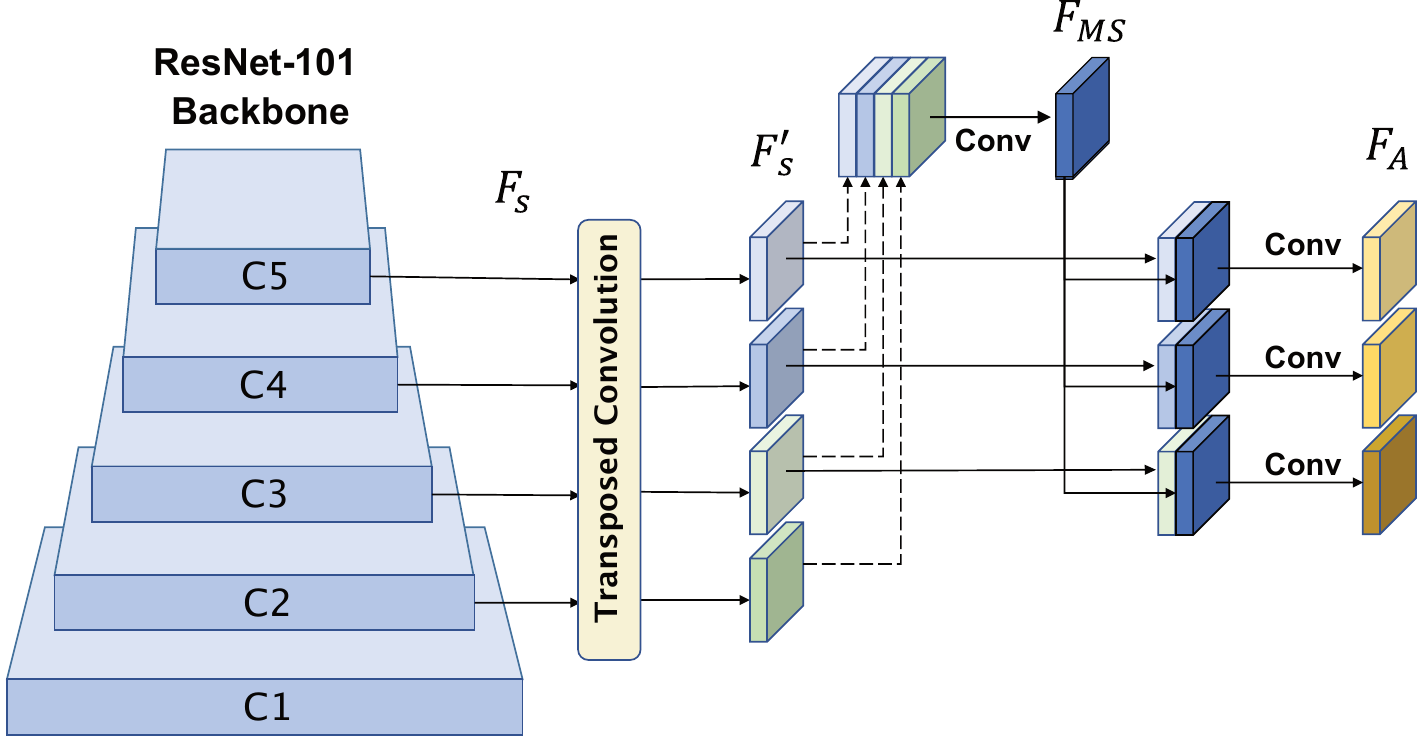}
    \caption{\textbf{Multi-scale feature fusion (MSFF) module.} Feature maps from the different scale feature maps of the backbone are combined to create aggregated features from all scales.}
    \label{fig:multiscale_fusion}
\end{figure}

To aggregate multi-scale features, while maintaining a high-resolution representation, we integrated a fusion module inspired by the method proposed in \citet{Deng_2018}. Considering the features at different scales indicated as $F_s$ where $s$ denote the scale level in the architecture (see Figure \ref{fig:multiscale_fusion}), features from each level $s$ are up-sampled through transposed convolution to the size of the highest resolution feature maps in the architecture, leading to enlarged feature maps $F'_s$. Next, all $F'_s$ are concatenated into a single tensor which is passed through a convolutional layer to integrate context from all scales into a single feature map $F_{MS} = conv([F'_0, F'_1, F'_2, F'_3, F'_4])$. In this manner, $F_{MS}$ encodes both low-level and high-level semantics learned at different stages. Finally, $F_{MS}$ is concatenated with each of the $F'_s$ feature maps and convolved to aggregate multi-scale information, creating multi-scale fused feature maps $F_A$.

Note that multi-scale feature fusion can be applied in the same way in both the backbone features and the FPN features. We opted to attach it on the backbone features as we believe it would create stronger representation which will get even further refined on the FPN.


\subsubsection{Attention mechanisms} 
\label{sec:attention_mechanisms}

We employed Criss-cross Attention Modules (CCAM) \cite{Huang_2020} and Convolutional Block Attention Modules (CBAM) \cite{Woo_2018}, specifically because of their fast and computationally efficient performance, which is paramount to introduce as less computational overhead as possible into the model, and therefore maintain low inference times. We attach the attention modules between the backbone and neck structures, as well as the neck and head of our network (see Figure \ref{fig:architecture}). The rationale behind the selection of these locations is that the addition of attention allows the model to extract richer context by aggregating local information with its corresponding global dependencies \cite{Huang_2020}. Additionally, attention aids to emphasize interdependent relationships between channel maps and between spatial regions without additional supervision \cite{Woo_2018}. Since the backbone network and the FPN are where most of the semantic context is distilled, it is a natural choice to attempt to refine their feature representations using attention, especially since mask prototypes and the prediction head benefit from better features.

\subsubsection{Loss function and anchor optimisation}

We use a combination of four losses: classification loss $L_{cls}$, bounding box regression loss $L_{box}$, mask loss $L_{mask}$, and semantic segmentation loss $L_{seg}$ with weights of 1, 1.5, 6.125 and 1, respectively. We employ softmax cross entropy for $L_{cls}$ with 1 positive label (instrument) and 1 background label. In the case of $L_{box}$ we used smooth-$L_1$ loss in the same way as \citet{Liu_2016}. $L_{mask}$ is defined as the pixel-wise binary cross entropy between the predicted masks and the ground truth. $L_{seg}$ is applied on layers that are only evaluated during training for additional feature richness~\citet{Bolya_2020}.


We further optimized the anchor boxes~\cite{Zlocha_2019} by defining 5 target scales and aspect ratios considering the size of our rescaled images ($600 \times 600$ pixels). The resulting scales were [0.435, 0.502, 0.578, 0.664, 0.762] and the resulting aspect ratios [0.267, 0.554, 1.0, 1.804, 3.746].


\section{Experiments and Results}{\label{sec:resultsandExperiments}}
\subsection{Data preparation}
As shown in Table \ref{tab:data_distribution}, the ROBUST-MIS dataset contains about 17\% empty frames (ef) on its training set. These frames do not have any visible instruments in them, and although we could have left them as negative examples for training, we opted to remove them from the training set. This decision was taken considering that the data already has plenty of negative examples in the frames’ background for the model to learn from; additionally, removing such frames speeds up the training. In the end, a total of 996 images with no visible instruments were discarded, leaving 4,987 frames in our training set. We then applied an 85-15\% split to the curated training set with 15\% for validation purposes. The training set was then randomly shuffled before creating the train and validation splits. We finally obtained a training set composed of 4,239 frames and a validation set comprised of 748 frames. As a final step, the training and validation datasets were converted to COCO-style for our instance segmentation framework, which involved extracting mask contours and generating bounding box coordinates from the provided annotation images and translating them to the target JSON format.

\subsection{Training setup}
\label{sec:training_setup}
The training was performed on an NVIDIA DGX-1 system consisting of 8 NVIDIA Volta-based GPUs; however, each model was trained on \emph{a single} GPU. The models were trained for up to 400,000 iterations with a learning rate of 0.001, momentum of 0.9, weight decay of $5e^{-4}$, and a batch size of 16. Input images were resized to $600 \times 600$ pixels during training, as this configuration constituted the best trade-off between GPU memory usage and segmentation performance. 
We applied data augmentation techniques to increase the model’s performance. These included random photometric distortions (i.e., changes in contrast, color-space, saturation, hue, brightness, and noise transformations) and affine transformations (i.e., random scaling, random sample crop, and random mirror).

%
\subsection{Ablation study setup}
\begin{table}[]
\fontsize{8}{9}\selectfont
\centering
\begin{tabular}{l|c|c|c}
\toprule
\multicolumn{1}{c|}{\begin{tabular}[c]{@{}c@{}}\bf{Model}\\ \bf{ Identifier}\end{tabular}} & \begin{tabular}[c]{@{}c@{}}\bf{ Attention}\\ \bf{Type}\end{tabular} & \begin{tabular}[c]{@{}c@{}}\bf{ Backbone}\\ \bf{Attention}\end{tabular} & \begin{tabular}[c]{@{}c@{}}\bf{FPN}\\ \bf{Attention}\end{tabular} \\ \hline
Base YOLACT++                                                                    &                                                          &                                                              &                                                         \\ \hline
CCAM-Backbone                                                                    & CCAM                                                     & \ding{51}                                                            &                                                         \\
CCAM-FPN                                                                         & CCAM                                                     &                                                              & \ding{51}                                                       \\
CCAM-Full                                                                        & CCAM                                                     & \ding{51}                                                            & \ding{51}                                                      \\ \hline
CBAM-Backbone                                                                    & CBAM                                                     & \ding{51}                                                            & \ding{51}                                                       \\
CBAM-FPN                                                                         & CBAM                                                     &                                                              & \ding{51}                                                       \\
CBAM-Full                                                                        & CBAM                                                     & \ding{51}                                                            & \ding{51}                                                       \\ \hline
\begin{tabular}[c]{@{}l@{}}CBAM-Full\\ + Aug\end{tabular}                        & CBAM                                                     & \ding{51}                                                            & \ding{51}                                                        \\ \hline
\begin{tabular}[c]{@{}l@{}}CBAM-Full\\ + Aug + Anch\end{tabular}                 & CBAM                                                     & \ding{51}                                                             & \ding{51}                                                        \\ \hline
\begin{tabular}[c]{@{}l@{}}CBAM-Full\\ + Aug + Anch + MSFF\end{tabular}            & CBAM                                                     & \ding{51}                                                             & \ding{51}                                                      \\
\bottomrule
\end{tabular}
\caption{Model configurations for our experiments with integration of different attention mechanisms in unique settings.}
\label{tab:attn_experiments}
\end{table}
Our experiments systematically integrate attention mechanisms in two strategic locations of the baseline YOLACT framework: 1) at the output of each convolutional block of the ResNet-101 backbone~\cite{He_2015}, and 2) at the multi-scale output features of the FPN~\cite{Lin_2017}. The incorporation of attention in these locations was alternated throughout experiments leading to three different network configurations:
\begin{enumerate}
    \item Exclusively incorporated in the backbone
    \item Exclusively incorporated in the FPN
    \item Integrated in the backbone and FPN simultaneously (which we refer to as a \emph{Full} configuration)
\end{enumerate}

\noindent At the end, six attention-based models were created by following this strategy, plus a baseline network without attention. Next, we selected the top performing configuration from the six mentioned network configurations. We then applied other optimization techniques to understand the network performance that included stronger data augmentation and anchor optimization. Finally, we added our multi-scale MSFF module. Table \ref{tab:attn_experiments} summarizes all the different model configurations.
%
\subsection{Metrics and assessment}
%
The algorithms' performance was evaluated following the guidelines defined by the ROBUST-MIS Challenge. Robustness performance was assessed considering the area-based metric multi-instance dice MI\_DSC and the distance-based multi-instance normalized surface dice MI\_NSD using the code implementations employed in the challenge which are provided in \citep{Ross_2019}. Furthermore, our reported model rankings were computed using the publicly available~\emph{challengeR} \cite{Wiesenfarth_2021} R package developed by the challenge organizers to accurately evaluate competitors. The ranking stability was investigated using bootstrapping for quantifying ranking variability using 1000 samples.

The robustness rankings are particularly focused on models’ capabilities in stage 3 of the challenge and pay particular attention on the worst-case performance of methods. For this reason, the robustness rankings are computed by aggregating the resulting scores for all the test cases by the 5\% percentile instead of by the mean or median.

We performed inference speed assessments by running inference tests on a 10 second video snippet from the ROBUST-MIS dataset a total of ten times per model. The reported frame rates were then aggregated by the mean. Inference was tested on a \emph{single} Tesla P100 GPU from the DGX-1 cluster with video multi-frame enabled.

\begin{table*}[t!]
\centering
\begin{tabular}{llccc}
\toprule
\multicolumn{1}{l|}{\bf Team/Model}                  & \multicolumn{1}{c|}{\bf Base Method}                                                                & \multicolumn{1}{c|}{\bf Aggr. MI\_DSC}  & \multicolumn{1}{c|}{\bf Aggr. MI\_NSD}  & \multicolumn{1}{c}{\bf FPS}         \\ \hline
\hline
\multicolumn{1}{l|}{\textit{www}}                & \multicolumn{1}{l|}{Mask R-CNN}                                                                 & \multicolumn{1}{c|}{0.31}           & \multicolumn{1}{c|}{0.35}           & \multicolumn{1}{c}{5*}          \\
\multicolumn{1}{l|}{\textit{Uniandes}}           & \multicolumn{1}{l|}{Mask R-CNN}                                                                 & \multicolumn{1}{c|}{0.26}           & \multicolumn{1}{c|}{0.29}           & \multicolumn{1}{c}{5*}          \\
\multicolumn{1}{l|}{\textit{SQUASH}}             & \multicolumn{1}{l|}{Mask R-CNN}                                                                 & \multicolumn{1}{c|}{0.22}           & \multicolumn{1}{c|}{0.26}           & \multicolumn{1}{c}{5*}          \\
\multicolumn{1}{l|}{\textit{CASIA\_SRL}}         & \multicolumn{1}{l|}{\begin{tabular}[c]{@{}l@{}}Dense Pyramid \\ Attention Network\end{tabular}} & \multicolumn{1}{c|}{0.19}           & \multicolumn{1}{c|}{0.27}           & \multicolumn{1}{c}{5*}          \\
\multicolumn{1}{l|}{\textit{fisensee}}           & \multicolumn{1}{l|}{2D U-Net}                                                                   & \multicolumn{1}{c|}{0.17}           & \multicolumn{1}{c|}{0.16}           & \multicolumn{1}{c}{12*}         \\
\multicolumn{1}{l|}{\textit{caresyntax}}         & \multicolumn{1}{l|}{Mask R-CNN}                                                                 & \multicolumn{1}{c|}{0.00}           & \multicolumn{1}{c|}{0.00}           & \multicolumn{1}{c}{5*}          \\
\multicolumn{1}{l|}{\textit{VIE}}                & \multicolumn{1}{l|}{Mask R-CNN}                                                                 & \multicolumn{1}{c|}{0.00}           & \multicolumn{1}{c|}{0.00}           & \multicolumn{1}{c}{5*}          \\ \hline
\multicolumn{1}{l|}{Base YOLACT++}               & \multicolumn{1}{l|}{YOLACT++}                                                                   & \multicolumn{1}{c|}{0.000}          & \multicolumn{1}{c|}{0.000}          & \multicolumn{1}{c}{75}          \\
\multicolumn{1}{l|}{CCAM-FPN}                    & \multicolumn{1}{l|}{YOLACT++}                                                                   & \multicolumn{1}{c|}{0.000}          & \multicolumn{1}{c|}{0.000}          & \multicolumn{1}{c}{60}          \\
\multicolumn{1}{l|}{CBAM-Backbone}               & \multicolumn{1}{l|}{YOLACT++}                                                                   & \multicolumn{1}{c|}{0.245}          & \multicolumn{1}{c|}{0.285}          & \multicolumn{1}{c}{65}          \\
\multicolumn{1}{l|}{CCAM-Full}                   & \multicolumn{1}{l|}{YOLACT++}                                                                   & \multicolumn{1}{c|}{0.308}          & \multicolumn{1}{c|}{0.333}          & \multicolumn{1}{c}{45}          \\
\multicolumn{1}{l|}{CCAM-Backbone}               & \multicolumn{1}{l|}{YOLACT++}                                                                   & \multicolumn{1}{c|}{0.313}          & \multicolumn{1}{c|}{0.338}          & \multicolumn{1}{c}{49}          \\
\multicolumn{1}{l|}{CBAM-FPN}                    & \multicolumn{1}{l|}{YOLACT++}                                                                   & \multicolumn{1}{c|}{0.315}          & \multicolumn{1}{c|}{0.333}          & \multicolumn{1}{c}{66}          \\
\multicolumn{1}{l|}{CBAM-Full}                   & \multicolumn{1}{l|}{YOLACT++}                                                                   & \multicolumn{1}{c|}{0.338}          & \multicolumn{1}{c|}{0.383}          & \multicolumn{1}{c}{65}          \\
\multicolumn{1}{l|}{CBAM-Full + Aug}             & \multicolumn{1}{l|}{YOLACT++}                                                                   & \multicolumn{1}{c|}{0.382}          & \multicolumn{1}{c|}{0.429}          & \multicolumn{1}{c}{63}          \\
\multicolumn{1}{l|}{CBAM-Full + Aug + Anch}      & \multicolumn{1}{l|}{YOLACT++}                                                                   & \multicolumn{1}{c|}{0.425}          & \multicolumn{1}{c|}{0.471}          & \multicolumn{1}{c}{\textbf{69}} \\
\multicolumn{1}{l|}{CBAM-Full + Aug + Anch + MS} & \multicolumn{1}{l|}{YOLACT++}                                                                   & \multicolumn{1}{c|}{\textbf{0.447}} & \multicolumn{1}{c|}{\textbf{0.489}} & \multicolumn{1}{c}{24}          \\ \bottomrule
\multicolumn{5}{l}{* Approximated from base method. Original measurement was not reported.} 
\end{tabular}
\caption{Evaluation results for stage 3 of the challenge. The upper part of the table shows the aggregated metrics for the competitors of the 2019 ROBUST-MIS challenge. The lower part of the table shows the results of the our developed models. MI\_DSC and MI\_NSD metrics are reported along with the base models and frame rates.}
\label{tab:metric_scores}
\end{table*}
\subsection{Results}
\label{sec:results}

In this section, we compare our results to the ROBUST-MIS challenge methods and present both quantitative and qualitative results.
\subsubsection{Quantitative results}

%
\begin{table*}[t!]
    \centering
\begin{tabular}{l|c|l|l|c|l|c}
\toprule
\multirow{2}{*}{\bf Model}& \multicolumn{2}{c|}{\textbf{Stage 1}}           & \multicolumn{2}{c|}{\textbf{Stage 2}}           & \multicolumn{2}{c}{\textbf{Stage 3}}           \\ 
 & MI\_DSC & \multicolumn{1}{c|}{MI\_NSD} & \multicolumn{1}{c|}{MI\_DSC} & MI\_NSD & \multicolumn{1}{c|}{MI\_DSC} & MI\_NSD \\ \hline \hline
Base YOLACT++               & 0.307   & 0.332                        & 0.000                        & 0.000   & 0.000                        & 0.000   \\ \hline
CCAM-Backbone               & 0.399   & 0.468                        & 0.317                        & 0.333   & 0.313                        & 0.338   \\ \hline
CCAM-FPN                    & 0.230   & 0.334                        & 0.000                        & 0.000   & 0.000                        & 0.000   \\ \hline
CCAM-Full                   & 0.403   & 0.447                        & 0.282                        & 0.333   & 0.308                        & 0.333   \\ \hline
CBAM-Backbone               & 0.425   & 0.462                        & 0.318                        & 0.350   & 0.245                        & 0.285   \\ \hline
CBAM-FPN                    & 0.433   & 0.485                        & 0.329                        & 0.365   & 0.315                        & 0.333   \\ \hline
CBAM-Full                   & 0.402   & 0.465                        & 0.331                        & 0.381   & 0.338                        & 0.383   \\ \hline
CBAM-Full + Aug             & 0.454   & 0.498                        & 0.313                        & 0.380   & 0.383                        & 0.430   \\ \hline
CBAM-Full + Aug + Anch      & {0.457}   & \bf{0.500}                        & \bf{0.439}                        &{0.468}   & 0.425                        & 0.471   \\ \hline
CBAM-Full + Aug + Anch + MS & \bf{0.460}   & {0.499}                        & {0.427}                        & \bf{0.478}   & \bf{0.447}                        & \textbf{0.487}   \\ \bottomrule
\end{tabular}
\caption{Metric comparison for all three test stages of ROBUST-MIS dataset. Best metric values are shown in bold for each test dataset.}
\label{tab:suppTab1}
\end{table*}
\begin{figure*}[t!]
    \centering
    \includegraphics[width=0.95\linewidth]{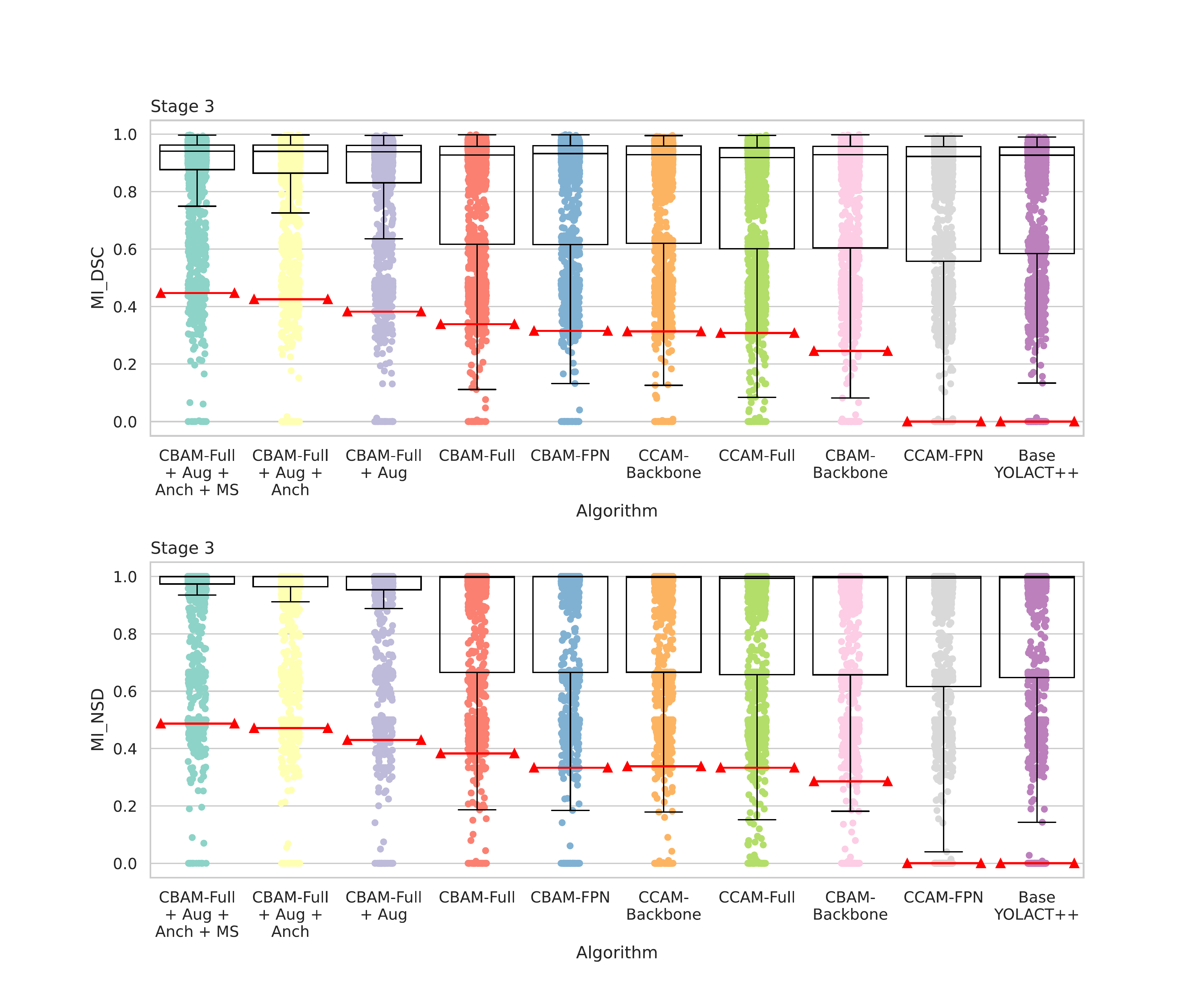}
    \caption{Dot-and-boxplots for the MI\_DSC and MI\_NSD showing the individual performances of algorithms on stage 3 of the challenge. The red lines indicate the value of the aggregated metric (by 5\% percentile) for each algorithm.}
    \label{fig:boxplots_stage1}
\end{figure*}

Table \ref{tab:metric_scores} shows the detailed result for the multi-instance segmentation task on the ROBUST-MIS dataset. It can be observed that overall, attention-based models show improvement over previous approaches for the aggregated MI\_DSC and MI\_NSD, and most notably FPS.
\emph{CCAM-Full}, \emph{CCAM-Backbone}, and \emph{CBAM-FPN} achieved competitive results in terms of MI\_DSC compared to the the top contestant \emph{www}, scoring 0.308, 0.313, and 0.315, respectively. On the other hand, the three models fall behind by 0.015 in average regarding MI\_NDS. Nonetheless, such a small difference in performance is out-weighted by the dramatic increase of inference speed of at least $9\times$ from all the models.
The highest metric scores from the initial ablation experiments correspond to \emph{CBAM-Full}, which resulted in 0.338 MI\_DSC and 0.383 MI\_NSD. \emph{CBAM-Full} presents a 2.8\% on MI\_DSC and 3.3\% on MI\_NSD improvement compared to the previously best model while attaining real-time inference speed of 65 FPS.
The additional data augmentation efforts applied in \emph{CBAM-Full + Aug} resulted in an improvement of 4.4\% and 4.6\% on MI\_DSC and on MI\_NSD, respectively, with respect to \emph{CBAM-Full}. \emph{CBAM-Full + Aug + Anch} greatly benefited from anchor optimization, which resulted in the model with the best balance between robustness and speed. It achieved scores of 0.425 MI\_DSC, 0.471 MI\_NSD, and runs at 69 FPS. For comparison, this model outperforms team \emph{www}'s by 11.5\% on MI\_DSC and 12.1\% on MI\_NSD while running $13.8\times$ faster.

Our most robust network and proposed architecture, namely \emph{CBAM-Full + Aug + Anch + MS}, reached 13.7\% MI\_DSC and 13.9\% MI\_NSD scores higher, compared to the top contestant of the challenge. It also outperforms \emph{CBAM-Full + Aug + Anch} by 2.2\% and 1.8\% margin. Nonetheless, \emph{CBAM-Full + Aug + Anch + MS}'s increased complexity has an impact on its inference speed which becomes evident when looking at its running speed of 24 FPS. Nevertheless, the model is still $4.8\times$ faster than the previous state-of-the-art. Table~\ref{tab:suppTab1} shows the results for each development stage of our final network on all three test datasets. It can be observed that for other Stage 1 and Stage 2 as well our proposed architecture with CBAM-Full + Aug + Anch showed improved performance over most combinations. However, in these cases the addition of the multi-scale feature fusion network (MS) only provided competitive result (e.g., 0.46 and 0.427 on MI\_DSC for Stage 1 and Stage 2, respectively). 

Figure \ref{fig:boxplots_stage1} shows the dot-and-boxplots of the MI\_DSC and MI\_NSD metric values obtained by each of our algorithms on Stage 3 test set used in the challenge. We can observe a large difference between the top model and the baseline model, as well as the progressive improvement from experiment to experiment. Despite most of the models are similar in terms of their median, the improvement is evident when looking at the aggregated metric values, as well as the first and third quartiles of the top model with our final model having the least deviations.

\begin{figure*}[t]
\centering
\subfloat[Qualitative improvements of worst performance frames from \emph{Base YOLACT++} to \emph{CBAM-Full}. Base model faults in alphabetical order: a) Missed instrument detection due to smoke overlay. b) False positive segmentation of bandage as instrument. c) Missed detection due to blood overlay. d) False positive segmentation of blood pool and shadow. e) Missed  detection due to motion blur. f) Missed detection of instrument covered in tissue.\label{fig:1a}]{\includegraphics[width=0.47\textwidth]{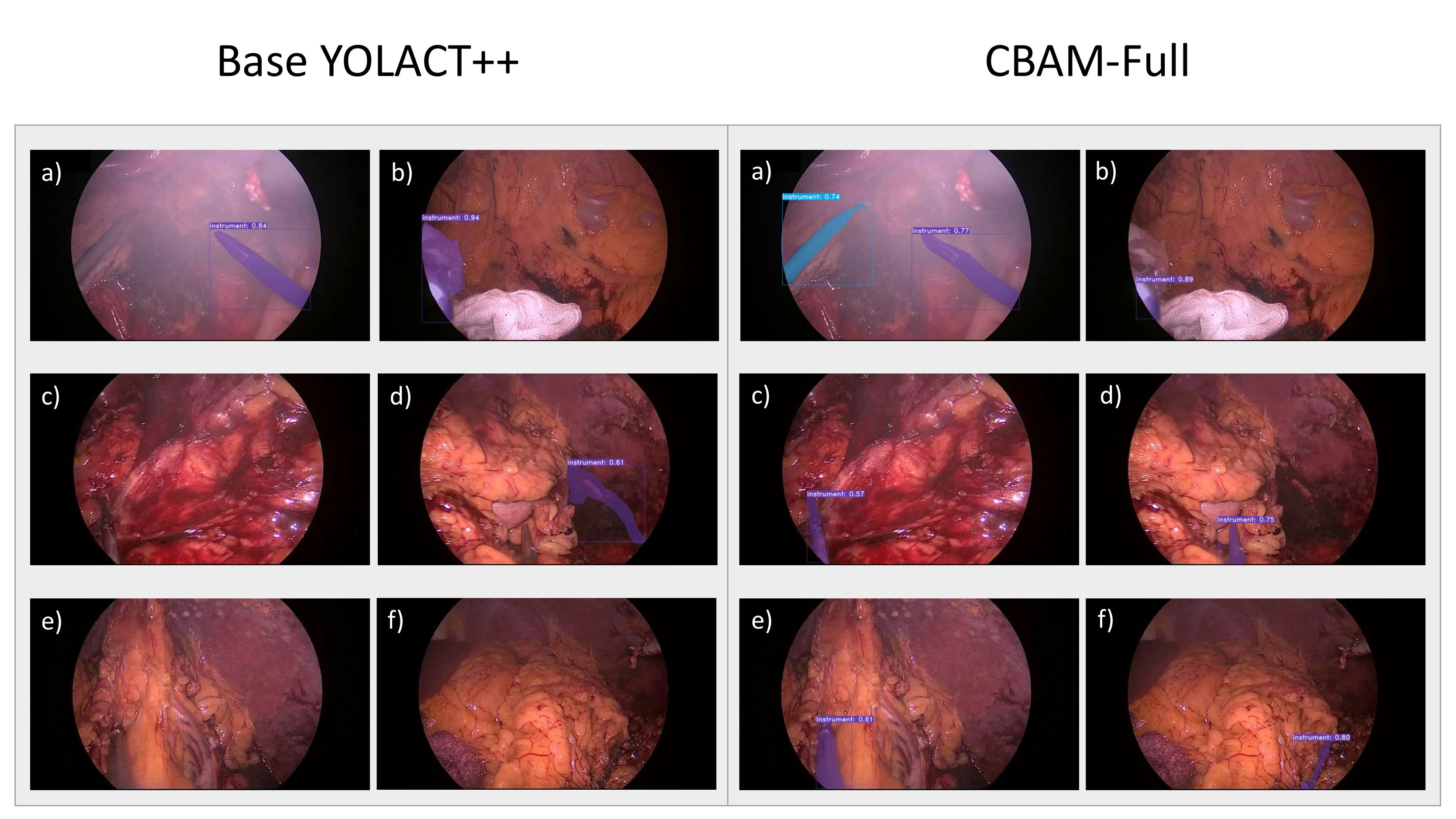}}\hfill
\subfloat[Qualitative improvements of worst performing frames from \emph{CBAM-Full} to \emph{CBAM-Full + Aug}. \emph{CBAM-Full} faults in alphabetical order: a) Missed vertical instrument. b) Single (split) instrument mistakenly regarded as two instances. c) Small instrument on the edge of the field of view. d) Missed vertical instrument and missed instrument on the edge of the field of view. e) Missed instruments on the edge of FOV. f) Missed transparent instrument. \label{fig:1b}] {\includegraphics[width=0.47\textwidth]{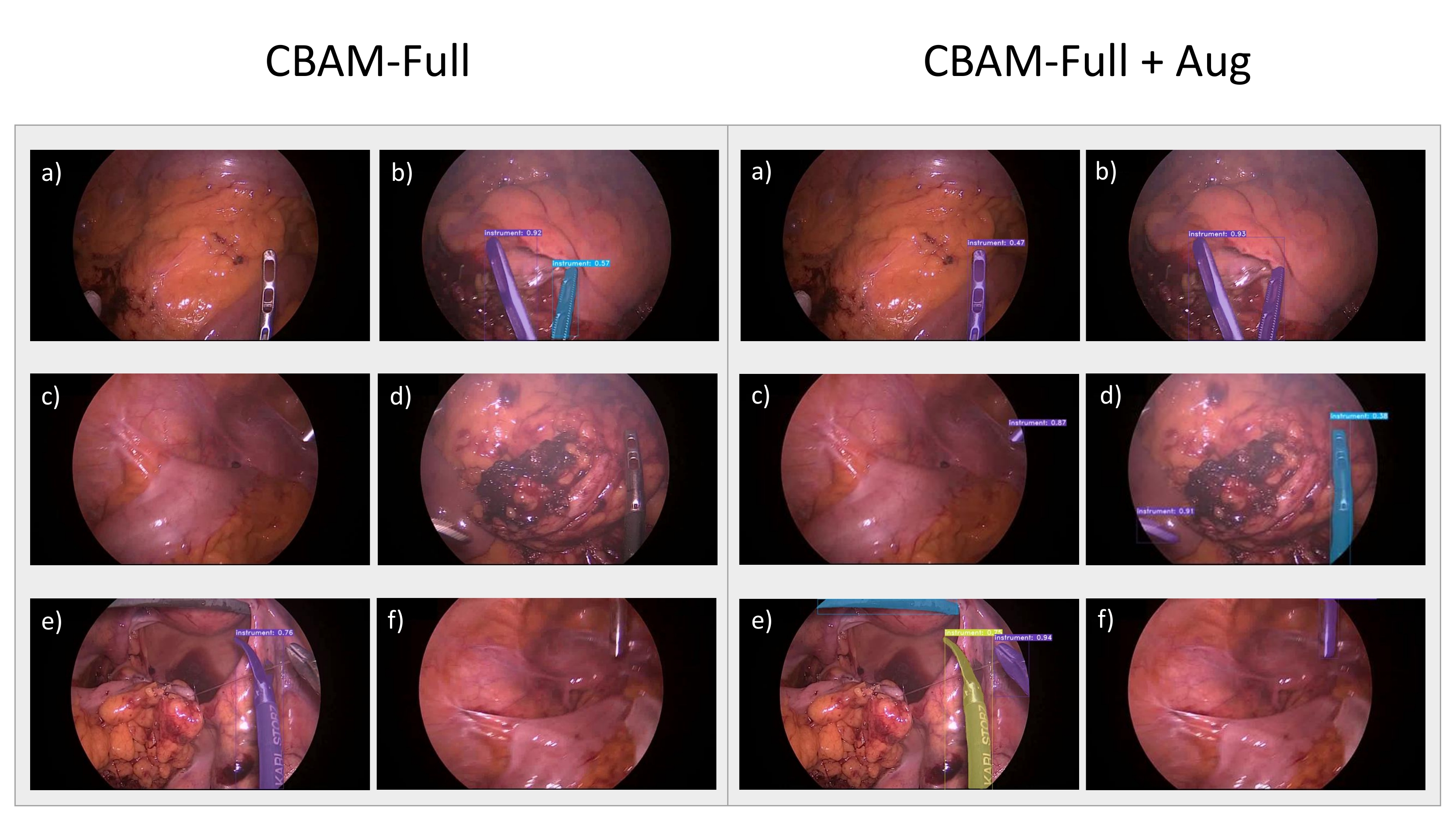}} \vfill
\subfloat[Qualitative improvements of worst performing frames from \emph{CBAM-Full + Aug} to \emph{CBAM-Full + Aug + Anch}. \emph{CBAM-Full + Aug} faults in alphabetical order: a) Missed vertical instrument. b) Missed transparent instrument. c) Missed vertical instrument on the edge of the field of view. d) Missed long horizontal instrument on the bottom of the image. e) Single instrument mistaken as two instruments and overlapped masks. f) Two missed detections of long, horizontal instruments.\label{fig:1c}]{\includegraphics[width=0.47\textwidth]{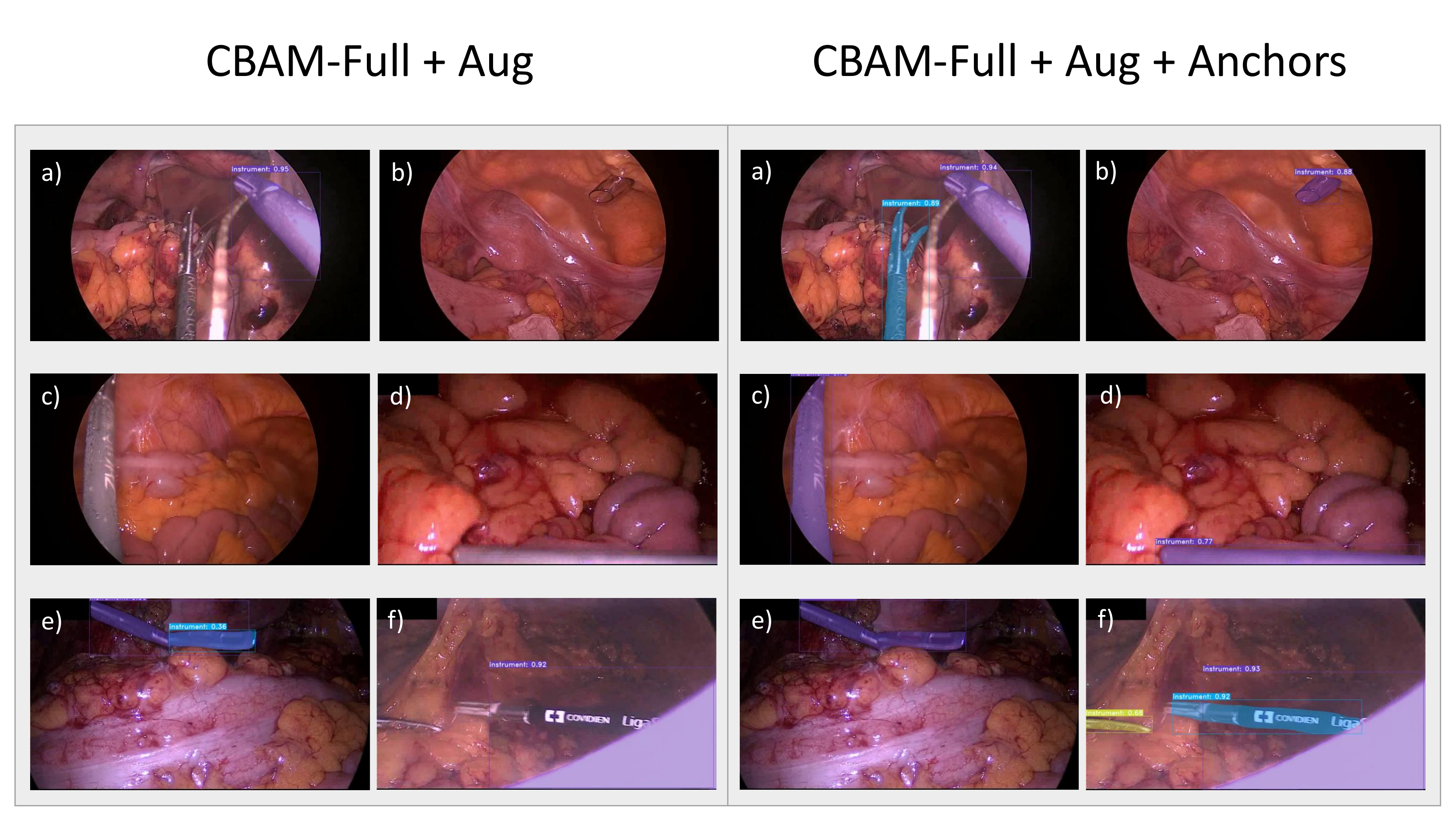}} \hfill
\subfloat[Qualitative improvements of worst performing frames from \emph{CBAM-Full + Aug + Anch} to \emph{CBAM-Full + Aug + Anch + MS}. \emph{CBAM-Full + Aug + Anch} faults in alphabetical order: a) Missed underexposed instrument. b) Missed transparent instrument. c) Missed blurred instruments. d) Missed blurred instrument due to reflection. e) Missed underexposed instrument. f) Missed detection due to motion artifacts.\label{fig:1d}]{\includegraphics[width=0.47\textwidth]{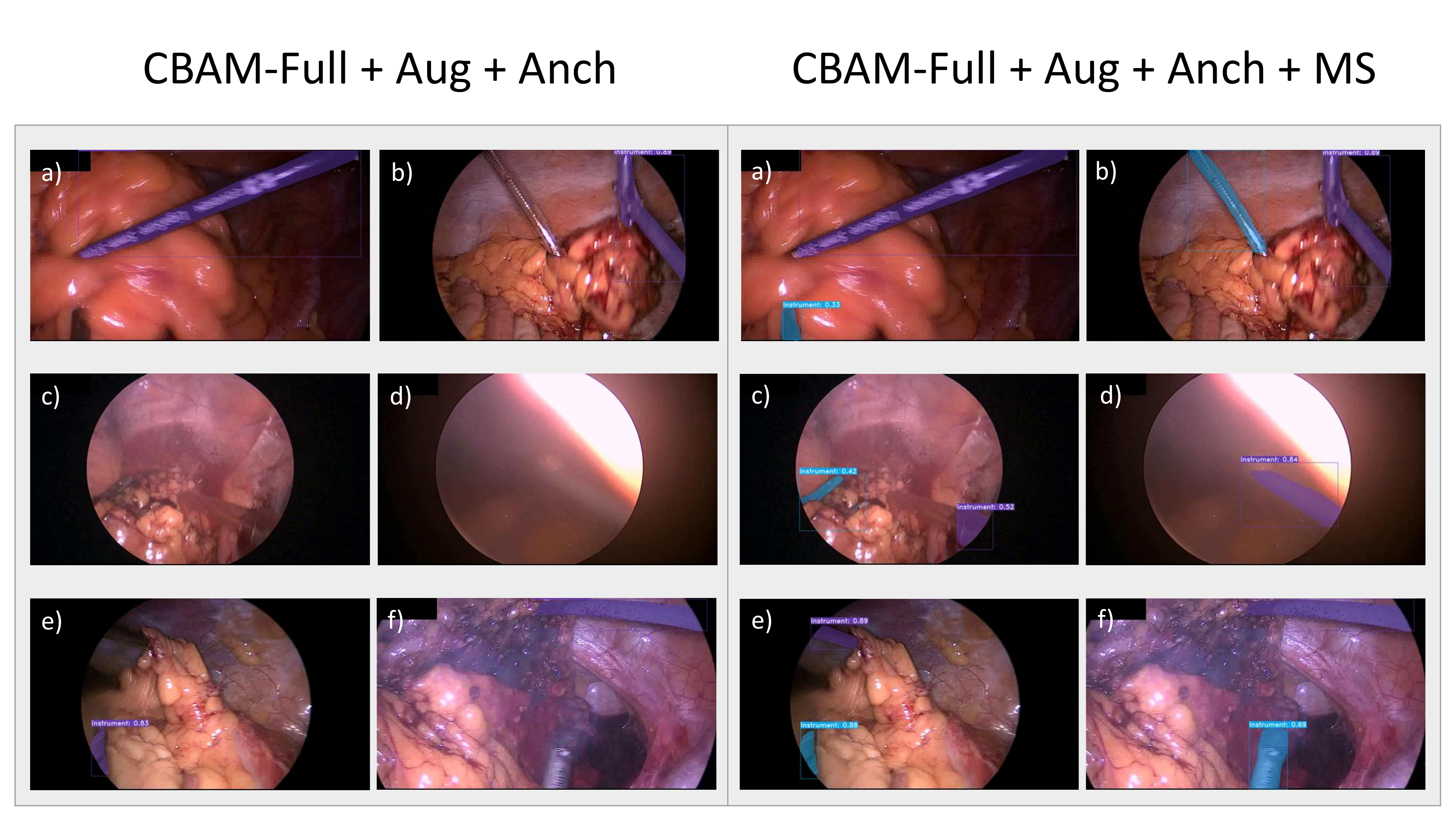}}
\caption{Qualitative comparison of several challenges instances by incremental improvements in our proposed model} \label{fig:total}
\end{figure*}

\subsubsection{Qualitative results}

To better understand the effects of the different components of our models, we performed a comparative analysis of the 100 frames with the worst performance of each network, as show in Figure \ref{fig:total}. \\ \\

Figure \ref{fig:1a} shows the qualitative comparison between worst frames of \emph{Base YOLACT++} and their corresponding frames from \emph{CBAM-Full}. We can observe from the images on the left side that the baseline model often presents missed detections, which hinders the model’s recall. Similarly, the model confuses tissue and other objects like bandages as instruments, evidencing its lack of robustness. On the other hand, \emph{CBAM-Full} overcomes some of these problems by attending to the important features, leading to better localization and segmentation results. We identified four different conditions that seemed particularly challenging for \emph{CBAM-Full}: transparent instruments, vertical instruments, small instruments on the edge of the field of view, and partially occluded instruments. According to \cite{Ross_2021}, most of these issues have also been challenging to previous participants. Figure \ref{fig:1b} illustrates the comparison of worst frames of \emph{CBAM-Full} and the improvements obtained after training with stronger data augmentation. 

In contrast to the baseline model with attention, we can observe that \emph{CBAM-Full + Aug} is capable of addressing problematic instances such as small instruments on the edge of the field of view. Similarly, transparent, partially occluded, and vertical instruments are now detected and segmented to a larger extent.

Howver, \emph{CBAM-Full + Aug} still presented recurrent missed detections on long vertical and transparent instruments. Figure \ref{fig:1c} illustrates the comparison of worst frames of \emph{CBAM-Full + Aug} and the improvements obtained after training with an optimized set of anchors. We can observe that the anchor optimization in \emph{CBAM-Full + Aug + Anch} led to additional detection and segmentation improvements not only on previously undetected objects but across all instruments. Figure \ref{fig:1d} illustrates the comparison of worst frames of \emph{CBAM-Full + Aug + Anch} and the improvements obtained after training with MSFF. 

We can observe that \emph{CBAM-Full + Aug + Anch + MS} is better at recognizing more challenging instances, like small under-exposed instruments, transparent instruments, and reflections.

\begin{figure}[t!]
\centering
 {\includegraphics[width=0.45\textwidth]{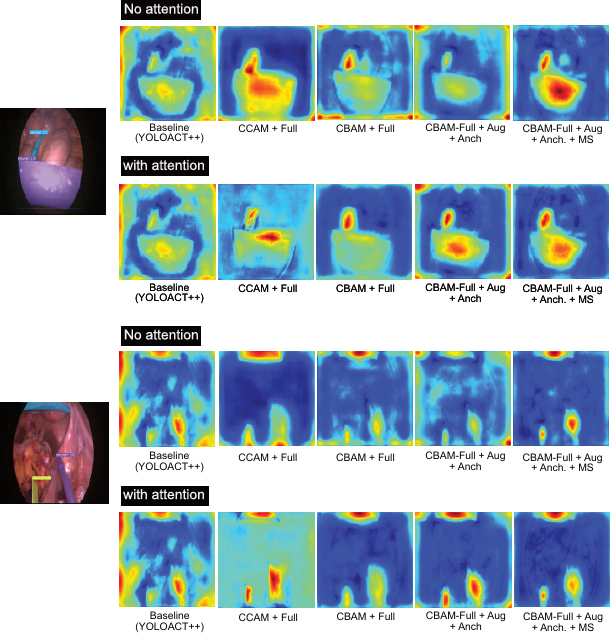}}
 \caption{\textbf{Network output with and without attention.} (top) Large part of image covered with instrument in purple and in blue (left). (bottom) Three instruments two at the bottom of image and one on the top of image (see left).  \label{fig:attention}}
\end{figure}

\section{Discussion}
\label{sec:discussion}
While deep learning has allowed us to design data driven solutions, generalisation remains a key issue that can cause significant performance degradation, especially in instrument segmentation applications where the intervention objects are of variable shapes and exposed to dynamic environments (e.g., smoke, flares, specularity etc, see Figure~\ref{fig:sampleimages}). In surgery, real-time performance of tool segmentation is of enormous importance for clinical utility. However, most methods build in the past relied on two stage networks that are not computationally efficient. 

To this end, we built over single stage YOLACT~\cite{yolact} and added several modifications to improve both the accuracy and robustness of our final surgical instrument segmentation algorithm. Among our models, we observed that those based on CBAM achieved slightly better performance that the ones based on CCAM. Regardless, attention-integrated models always outperformed the attention-less baseline in terms of robustness. Figure~\ref{fig:boxplots_stage1} shows dot-and-boxplots of the metric values obtained by each algorithm over all test cases in stage 3 dataset (unseen test dataset in the MIS challenge). 

We can observe that adding attention mechanisms boosts the performance compared to the baseline model used in our architecture design~\cite{yolact}. This especially true for instances below the third quartile, which are the most important for our performance metrics. Among the three model variations to which we added CCAM attention modules, CCAM-Backbone achieved the best results in terms of robustness (0.313 MI\_DSC and 0.338 MI\_NSD, see Table~\ref{tab:metric_scores}). This indicates that the contextually enriched feature maps from the ResNet-101 backbone are powerful enough to generate more accurate mask prototypes and coefficients in the YOLACT architecture, and ultimately better segmentation outputs. However, the CBAM-Full model outperformed all other attention modules by nearly 9\% on MI\_DSC over the best performing CCAM-Backbone. 

The superiority of CBAM when integrated together with backbone and FPN all together enhances both channel-wise and block attentions better to represent local and global features well. The FPN layer allows to capture size variability present in the dataset for which CCAM yielded zero on aggregated MI\_DSC score (Figure~\ref{fig:boxplots_stage1}). 

We thoroughly investigated on model improvement for generalisability through different mechanisms such as data augmentation and optimisation of anchor weights. These steps were experimentally proven to be right directions giving subsequent increase in both aggregated MI\_DSC and MI\_NSD (see Table~\ref{tab:metric_scores}, Figure~\ref{fig:1a} and Figure~\ref{fig:1b}). For example, by using augmentation, we observed that the images with different view points (mostly long sized instruments either straight or slightly oblique) performed better; this can be because of low number of such samples in the dataset (Figure~\ref{fig:1b}). Similarly, for the variable size instruments optimising anchors provided substantial improvements (Figure~\ref{fig:1c}). We also noticed issues with the small instrument appearing either on the bottom or sides of the frame, and also for those which appeared as background, due to specularity or covering large tissue area with similar color (see Figure~\ref{fig:1c}, left). The use of the Multi-scale feature fusion (MSFF) network allowed us to fuse features at different scales and layers (Figure~\ref{fig:1d}). 

Our architecture with MSFF network integrated together with attention maps allowed to transfer both local and global context fusing high and low-level feature representations. As a result, we observed further improvement of nearly 2\% on aggregated MI\_DSC above the CBAM-Full with augmentation and anchor approach (Table~\ref{tab:metric_scores}). Our final best approach has near real-time performance of 24 FPS on NVIDIA Volta GPU. However, our second best performing method can run at 69 FPS which is above the required real-time performance in most cases (i.e., 45 FPS). A similar, performance improvement was observed for other two test datasets (Table~\ref{tab:suppTab1}).

To better understand the behavior of the two types of attention modules implemented in our experiments, we visualize the attention maps generated by 3 different models, in addition to the feature maps of the baseline model (see Figure \ref{fig:attention}). We chose the \emph{CBAM-Full}, \emph{CCAM-Full}, and \emph{CBAM-Full + Aug + Anch + MS} configurations for visualization purposes as they show the refined feature maps using the two different attentions modules. 
Figure \ref{fig:attention} illustrate the activation maps taken before and after the attention modules. We visualize the effects of attention on this specific point of the network since mask prototypes and coefficients are extracted from FPN features and backbone network (Full). We can observe that over-mixing problem of \emph{CCAM-Full}. In fact, activation maps before CCAM in the Full model are cleaner and more discriminative than the actual attention output. This confirms that CCAM excels at refining features that have not been mixed before.
On the other hand, CBAM-based models produce slightly lower quality feature maps from attention modules in the backbone, as it can be observed by the clouds and blobs on the top maps of the figure. Nonetheless, the features are drastically refined after passing the attention modules in the Full, leading to clean and discriminative activations. This corroborates our hypothesis that CBAM is better at refining previously aggregated data, yielding a superior overall performance. 

Furthermore, our current work is built upon publicly available retrospective dataset providing a strong evidence of robustness ability compared to presented methods at the MIS challenge and current literature. In future work, we will validate on prospective data and benchmark it in clinical settings.

\section{Conclusion}
\label{sec:conclusion}
We have developed a real-time novel architecture that builds upon a single-stage instance segmentation method. Through our step-wise solution to different problems in surgical data, we have identified and integrated components that can be deal with existing challenges for robust segmentation of surgical instruments. We have provided comprehensive experiments and analysis that supports our final architecture development and its impact on surgical tool segmentation in clinic. Our method outperformed all methods reported in recently conducted ROBUST-MIS challenge.

\section*{Acknowledgements}
\noindent

The authors would like to thank the AI Hub  and the CIIOT at Tecnologico de Monterrey for their support for carrying the experiments on their NVIDIA's DGX computer.

\bibliographystyle{cas-model2-names}

\bibliography{cas-refs}

\end{document}